\documentclass[prd,showpacs]{revtex4}
\usepackage{amsmath,amssymb}

\begin{document}
\title{Magnetic Casimir effect of a Lorentz-violating scalar with higher order derivatives}
\author{Andrea Erdas}
\email{aerdas@loyola.edu}
\affiliation{Department of Physics, Loyola University Maryland, 4501 North Charles Street,
Baltimore, Maryland 21210, USA}
%\date{March, 2025}
\begin {abstract} 
In this paper I study the Casimir effect caused by a charged and massive scalar field that breaks Lorentz invariance in a CPT-even, aether-like manner. The breaking of Lorentz invariance is implemented by a constant space-like vector directly coupled to higher order derivatives of the field. I take this vector to be space-like to avoid non-causality problems that could arise with a time-like vector. I examine the two scenarios of the scalar field satisfying either Dirichlet or mixed boundary conditions on a pair of plane parallel plates. I use the zeta function technique to investigate the effect of a constant magnetic field, perpendicular to the plates, on the Casimir energy and pressure. I  examine two different directions of the unit vector: parallel and perpendicular to the plates. I fully examine both scenarios for both types of boundary conditions and, in both cases and for both types of boundary conditions, I obtain simple analytic expressions of the Casimir energy and pressure in the three asymptotic limits of strong magnetic field, large mass, and small plate distance.
\end {abstract}
\maketitle
%%%%%%%%%%%%%%%%%%%%%%%%%%%%%%%%%%%%%%%%%%%%%%%%%%%%%%%%%%%%%
\section{Introduction}
\label{1}
Seventy seven years ago Hendrik Casimir made the first theoretical prediction  of an attractive force between two uncharged and conducting parallel plates in vacuum, entirely due to quantum effects \cite{Casimir:1948dh}. This prediction was confirmed by experiments ten years later  \cite{Sparnaay:1958wg}, and many and increasingly more accurate experimental verifications followed throughout the decades \cite{Bordag:2001qi,Bordag:2009zz}. The Casimir effect is strongly dependent on the boundary conditions at the plates of the quantum field responsible of the Casimir force. Dirichlet or Neumann boundary conditions cause an attraction between the plates, mixed (Dirichlet-Neumann) boundary conditions cause repulsion \cite{Boyer:1974}. 

While standard quantum field theory strictly prohibits the violation of Lorentz invariance, newer theories propose models where Lorentz violation leads to space-time anisotropy \cite{Ferrari:2010dj,Ulion:2015kjx}. Lorentz symmetry breaking mechanisms have been proposed in some quantum gravity theories \cite{Alfaro:1999wd,Alfaro:2001rb}, in models that propose variation of some coupling constants \cite{Kostelecky:2002ca,Anchordoqui:2003ij,Bertolami:1997iy}, and in string theory \cite{Kostelecky:1988zi}, where some vector and tensor field components could have non-vanishing expectation values which, in turn, lead to a spontaneous Lorentz symmetry breaking  at the Planck energy scale. A detailed list of papers that study various consequences of Lorentz symmetry breaking is available in Refs. \cite{Cruz:2017kfo,ADantas:2023wxd}. 
Implications of the existence of Lorentz violation in the Casimir effect have been studied in the case of Lorentz-breaking extensions of QED \cite{Frank:2006ww,Kharlanov:2009pv,Martin-Ruiz:2016lyy}. More recently, a few papers have examined the case of a real scalar field  in vacuum \cite{Cruz:2017kfo,ADantas:2023wxd} and in a medium at finite temperature \cite{Cruz:2018bqt}, or a complex and charged scalar field in vacuum in the presence of a magnetic field \cite{Erdas:2020ilo}, and in a medium at finite temperature with a magnetic field \cite{Erdas:2021xvv}. 
These more recent papers investigate a modified Klein-Gordon model that breaks Lorentz symmetry in a CPT-even, aether-like manner. While most of these papers investigate a scenario where the Lorentz violation is implemented by the presence of a constant vector directly coupled to the lowest order derivatives of the scalar field, one of them \cite{ADantas:2023wxd}, investigates the scenario where a constant space-like vector is directly coupled to higher order derivatives of the field.

Several authors have studied the magnetic Casimir effect in Lorentz symmetric spacetime \cite{CougoPinto:1998td,CougoPinto:1998jg,Erdas:2013jga,Erdas:2013dha}, or in spacetime where Lorentz violation is implemented by a constant vector coupled to the lowest order derivatives of the field \cite{Erdas:2020ilo,Erdas:2021xvv,Droguett:2025frq}, but there has not been a study of the magnetic Casimir effect of a charged scalar field that breaks the Lorentz symmetry in a CPT-even, aether-like manner implemented by a unit space-like vector coupled to arbitrarily high order derivatives of the field. This work intends to fill that gap and provide theoretical predictions of the magnetic field effects on the quantum vacuum of the modified Klein-Gordon model introduced in Ref. \cite{ADantas:2023wxd}.
In this paper I will investigate the effect of a uniform magnetic field on the Casimir energy and pressure due to a Lorentz-violating scalar field, by studying a model similar to the one first presented in Ref.  \cite{ADantas:2023wxd}: a charged scalar field that breaks Lorentz symmetry and satisfies either Dirichlet or mixed boundary conditions on a pair of large parallel plates. I will not examine the case where the field obeys Neumann boundary conditions on the plates, because it produces the same results obtained with Dirichlet boundary conditions.

In Sec. \ref{2} of this paper I present the theoretical model of a charged scalar field that breaks Lorentz symmetry in an aether-like and CPT-even manner, by way of the coupling of a space-like unit vector to arbitrarily high derivative of the field, and use the zeta function technique \cite{Elizalde:1988rh,Elizalde:2007du} to obtain an expression of the vacuum energy without and with magnetic field, containing integrals and infinite sums. In Sec. \ref{3} I examine the case of a Lorentz asymmetry in the direction parallel to the Casimir plates and calculate the Casimir energy, obtaining simple analytic expressions for the energy in the short plate distance limit, large magnetic field limit, and large mass limit. In Sec. \ref{4} I investigate the case of Lorentz 
anisotropy perpendicular to the plates, calculate the Casimir energy, and obtain simple expressions for it in the three limits listed above. In Sec. \ref{5} I calculate the Casimir pressure for all the cases described above. In all these sections I focus first on the scalar field obeying Dirichlet boundary conditions at the plates, and then on it obeying mixed boundary conditions at the plates. My conclusions, along with a detailed discussion of my results are presented in Sec. \ref{6}.
%%%%%%%%%%%%%%%%%%%%%%%%%%%%%%%%%%%%%%%%%%%%%%%%%%%%%%%%%%%%%
\section{The model}
\label{2}

In this work, I investigate the Casimir effect due to a charged scalar field $\phi$ of mass $m$ that breaks the Lorentz symmetry in an aether-like and CPT-even  manner. The Lorentz-symmetry breaking is implemented by a constant space-like unit vector $u^\mu$ coupled to higher order derivatives of the field, as presented in the theoretical model introduced by Ref. \cite{ADantas:2023wxd}. The modified Klein Gordon equation for this field is
\begin{equation}
[\Box 
+l^{2(\epsilon -1)}(u\cdot\partial)^{2\epsilon}+m^2]\phi=0,
\label{KG}
\end{equation}
where the unit vector $u^\mu$ points in the direction in which the Lorentz symmetry is broken, the length $l$ is of the order of the inverse of the energy scale at which the Lorentz symmetry is broken, and the parameter $\epsilon$ is a positive integer that I call the critical exponent, in analogy to what is done for the fermion field case in Horava-Lifshitz theories \cite{daSilva:2019iwn,Farias:2011aa,Erdas:2023wzy}. The case with $\epsilon=1$ has been studied in Ref. \cite{Cruz:2017kfo} and, for the magnetic case, in \cite{Erdas:2020ilo}. Here I focus on $\epsilon\ge 2$ and, to avoid non-causality problems, I take $u^\mu$ to be spacelike. My aim is to study how this type of space-time anisotropy and the presence of a magnetic field modify the Casimir effect. I consider two square plates of side $L$ perpendicular to the $z$ axis, and a constant magnetic field $\vec B$ pointing in the $z$ direction. The two plates are located at $z=0$ and $z=a$. 
I will use the zeta function technique to study this problem, and investigate Dirichlet and mixed boundary conditions of the field $\phi$ at the plates. Investigating Neumann boundary conditions is trivial, since it produces the same results found with Dirichlet boundary conditions. I will study the cases when the unit four-vector $u^\mu$ is parallel to the plates, and perpendicular to the plates.

At this initial stage I do not include the magnetic field, but will introduce it later. I start by examining the case of $u^\mu$ parallel to the plates
\begin{equation}
u^\mu=\left(0,{1\over \sqrt{2}},{1\over \sqrt{2}},0\right),
\label{upar}
\end{equation}
and obtain the following dispersion relation for $\phi$
\begin{equation}
\omega^2_{{\bf k},n}=k^2_x+k^2_y+l^{2(\epsilon - 1)}(-1)^\epsilon\left({k^{2\epsilon}_x+k^{2\epsilon}_y\over2}\right)+k^2_z+m^2,
\label{dispersion_1}
\end{equation}
where $k_x$ and $k_y$ can take any real value and $k_z$ takes only discrete values. For Dirichlet boundary conditions, 
\begin{equation}
k_z={n\pi\over a},
\label{kz_D}
\end{equation}
with $n=1,2,3,\cdots$, while, for mixed boundary conditions
\begin{equation}
k_z=\left(n+\frac{1}{2}\right)\frac{\pi}{ a},
\label{kz_mixed}
\end{equation}
with $n=0,1,2,3,\cdots$. The vacuum expectation value of the hamiltonian is the vacuum energy
\begin{equation}
<0|\hat{H}|0> = E_0=\left(L\over 2\pi\right)^2\int d^2{\bf k} \sum_{n}\omega_{{\bf k},n}.
\label{E0_1}
\end{equation}
I do a change of integration variables from cartesian to plane polar coordinates, $(k_x, k_y) \rightarrow (k, \theta)$, and then another change of variable to $u=ak$, to obtain
\begin{equation}
E_0={L^2\over 4\pi^2a^3}\int_0^\infty udu \int_0^{2\pi} d\theta \sum_{n}\left[u^2+a^2k^2_z+m^2a^2+\left(l\over a\right)^{2(\epsilon - 1)}(-1)^\epsilon u^{2\epsilon}\left({\cos^{2\epsilon}\theta+\sin^{2\epsilon}\theta\over 2}
\right)\right]^{1\over 2},
\label{E0_2}
\end{equation}
where $k_z$ is given by Eq. (\ref{kz_D}) for Dirichlet boundary conditions, and by Eq. (\ref{kz_mixed}) for mixed boundary conditions, and the sum is over the discrete values of $k_z$. Since the dimensionless parameter ${l\over a}\ll 1$, I expand up to first order in this small parameter and find
\begin{equation}
E_0={L^2\over 4\pi^2a^3}\int_0^\infty \!\!\!udu \int_0^{2\pi} \!\!d\theta \sum_{n}\left\{ \left[u^2+a^2k^2_z+m^2a^2\right]^{1\over 2}
+{(-1)^\epsilon\over 4}\left(l\over a\right)^{2(\epsilon - 1)}\!\! (\cos^{2\epsilon}\theta+\sin^{2\epsilon}\theta)u^{2\epsilon}
\left[u^2+a^2k^2_z+m^2a^2\right]^{-{1\over 2}}\right\},
\label{E0_3}
\end{equation}
where the first term is the vacuum energy in the absence of Lorentz violation and the second term, ${\tilde E}_0$, is the Lorentz violating correction to the vacuum energy
\begin{equation}
{\tilde E}_0={L^2\over 4\pi^2a^3}\left(l\over a\right)^{2(\epsilon - 1)}\!\!{(-1)^\epsilon\over 4} 
\int_0^\infty udu \int_0^{2\pi} d\theta \sum_{n}\
u^{2\epsilon}\left({\cos^{2\epsilon}\theta+\sin^{2\epsilon}\theta}\right)
\left[u^2+a^2k^2_z+m^2a^2\right]^{-{1\over 2}}.
\label{E0t_01}
\end{equation}
Using
\begin{equation}
 \int_0^{2\pi} d\theta \left({\cos^{2\epsilon}\theta+\sin^{2\epsilon}\theta}\right)= 4\pi{(2\epsilon -1)!!\over(2\epsilon)!!},
\label{theta_1}
\end{equation}
I find
\begin{equation}
{\tilde E}_0={L^2\over 4\pi a^3}\left(l\over a\right)^{2(\epsilon - 1)}\!\!{(-1)^\epsilon} 
{(2\epsilon -1)!!\over(2\epsilon)!!}
\int_0^\infty udu  \sum_{n}
u^{2\epsilon}
\left[u^2+a^2k^2_z+m^2a^2\right]^{-{1\over 2}}.
\label{E0t_02}
\end{equation}

Next, I examine the scenario where the unit vector $u^\mu$ is perpendicular to the plates
\begin{equation}
u^\mu=\left(0,0,0,1\right),
\label{uperp}
\end{equation}
obtaining the following dispersion relation for $\phi$
\begin{equation}
\omega^2_{{\bf k},n}=k^2_x+k^2_y+k^2_z+l^{2(\epsilon - 1)}(-1)^\epsilon k^2_z+m^2.
\label{dispersion_2}
\end{equation}
I proceed as I did above and do a change of variables from cartesian to polar coordinates, and then a second change of variable to $u=ka$, to obtain
\begin{equation}
E_0={L^2\over 2\pi a^3}\int_0^\infty udu  \sum_{n}\left[u^2+a^2k^2_z+m^2a^2+\left(l\over a\right)^{2(\epsilon - 1)}(-1)^\epsilon (ak_z)^{2\epsilon}\right]^{1\over 2},
\label{E0_4}
\end{equation}
where I did also the straightforward angular integration. I expand in the small parameter $l\over a$ and find
\begin{equation}
E_0={L^2\over 2\pi a^3}\int_0^\infty udu  \sum_{n}\left\{ \left[u^2+a^2k^2_z+m^2a^2\right]^{1\over 2}
+{1\over 2}\left(l\over a\right)^{2(\epsilon - 1)}\!\!(-1)^\epsilon (ak_z)^{2\epsilon}
\left[u^2+a^2k^2_z+m^2a^2\right]^{-{1\over 2}}\right\},
\label{E0_5}
\end{equation}
where the first term is the vacuum energy without Lorentz violation as seen in Eq. (\ref{E0_3}), and the second term is the Lorentz violating correction to the vacuum energy
\begin{equation}
{\tilde E}_0={L^2\over 4\pi a^3}\left(l\over a\right)^{2(\epsilon - 1)}\!\!{(-1)^\epsilon} 
\int_0^\infty udu  \sum_{n}
(ak_z)^{2\epsilon}
\left[u^2+a^2k^2_z+m^2a^2\right]^{-{1\over 2}}.
\label{E0t_03}
\end{equation}

Now I introduce the magnetic field and, as it is shown in Refs. \cite{CougoPinto:1998jg,Erdas:2013jga,Erdas:2013dha,Erdas:2020ilo}, its presence modifies the vacuum energy $E_0$ by changing $k^2_x+k^2_y=k^2$ into $(2\ell+{1})eB$ and $\int_0^\infty k dk$ into
$eB\sum\limits_{\substack{\ell =0}}^\infty$, where $e$ is the charge of the scalar field and $\ell=0,1,2,\cdots$, labels the Landau levels. Since I am using the variable $u=ak$, I need to make the following two replacements into Eq. (\ref{E0_3}), $u^2\rightarrow a^2eB(2\ell+1)$ and $\int_0^\infty u du \rightarrow a^2eB\sum\limits_{\substack{\ell =0}}^\infty$, to obtain the vacuum energy $E_0$ in the presence of a magnetic field $B$ perpendicular to the plates when the Lorentz violating vector $u^\mu$ is parallel to the plates
\begin{equation}
E_0={L^2eB\over 2\pi a}  \sum_{n,\ell}\left\{a \left[eB(2\ell+1)+k^2_z+m^2\right]^{1\over 2}
+{(-1)^\epsilon\over 2a}{(2\epsilon -1)!!\over(2\epsilon)!!}\left(l\over a\right)^{2(\epsilon - 1)}\!\!\! [a^2eB(2\ell+1)]^{\epsilon}
\left[eB(2\ell+1)+k^2_z+m^2\right]^{-{1\over 2}}\right\},
\label{EB_1}
\end{equation}
where I also did the angular integration.

When $u^\mu$ is perpendicular to the plates, I make the two aforementioned substitutions to
Eq. (\ref{E0_5}) and obtain the vacuum energy in a magnetic field 
\begin{equation}
E_0={L^2eB\over 2\pi a } \sum_{n,\ell}\left\{a \left[eB(2\ell+1)+k^2_z+m^2\right]^{1\over 2}
+{(-1)^\epsilon\over 2a}\left(l\over a\right)^{2(\epsilon - 1)}\!\! (ak_z)^{2\epsilon}
\left[eB(2\ell+1)+k^2_z+m^2\right]^{-{1\over 2}}\right\}.
\label{EB_2}
\end{equation}
Notice that the first term in the last two equations is identical and represents the magnetic Casimir energy without Lorentz violation, while the second terms are its Lorentz violating correction in the two scenarios examined in this work. 

In the next sections I will start from Eqs. (\ref{EB_1}) and (\ref{EB_2}) and use the zeta function technique to obtain the Lorentz violating corrections to the vacuum energy for the cases of $u^\mu$ parallel to the plates and $u^\mu$ perpendicular to the plates. I will obtain these energy corrections for Dirichlet and mixed boundary conditions.
%%%%%%%%%%%%%%%%%%%%%%%%%%%%%%%%%%%%%%%%%%%%%%%%%%%%%%%%%%%%%%%%%%%%%
\section{Casimir energy for $u^\mu$ parallel to the plates}
\label{3}
I use the identity
\begin{equation}
z^{-s}={1\over \Gamma(s)}\int_0^\infty  t^{s-1}e^{-zt} dt,
\label{z}
\end{equation}
where $\Gamma(s)$ is the Euler gamma function, to rewrite Eq. (\ref{EB_1}) as
\begin{equation}
E_0=E_0'+{\tilde E}_0,
\label{E0_6}
\end{equation}
where $E_0'$, the magnetic Casimir energy in absence of Lorentz violation, is
\begin{equation}
E_0'={L^2eB\over 2\pi a } \sum_{n}{1\over \Gamma(-{1\over 2})}\int_0^\infty  t^{-3/2}{e^{-(k^2_z+m^2)a^2t}\over 2\sinh(eBa^2t)}dt,
\label{E0p_01}
\end{equation}
and ${\tilde E}_0$, the Lorentz violating correction to the magnetic Casimir energy, is
\begin{equation}
{\tilde E}_0={L^2eB\over 2\pi a } 
{(-1)^\epsilon\over 2}{(2\epsilon -1)!!\over(2\epsilon)!!}\left(l\over a\right)^{2(\epsilon - 1)}\!\! \sum_{n,\ell} [a^2eB(2\ell+1)]^{\epsilon}{1\over \Gamma({1\over 2})}\int_0^\infty  t^{-1/2}e^{-[eB(2\ell+1)+k^2_z+m^2]a^2t}dt.
\label{E0t_04}
\end{equation}
Notice that I used
\begin{equation}
\sum_{\ell=0}^\infty e^{-(2\ell +1)z}={1\over 2 \sinh z},
\label{sinh}
\end{equation}
in the equation for $E'_0$. The Casimir energy in magnetic field and without Lorentz violation shown in Eq. (\ref{E0p_01}) has been obtained in the past by several authors, see, for example, Refs. \cite{CougoPinto:1998td,Erdas:2013jga,Erdas:2013dha,Erdas:2020ilo} and references within, for Dirichlet and mixed boundary conditions. My result for $E_0'$ is in full agreement with all papers in the literature and therefore, if needed, I will quote their result.

In order to proceed and evaluate Eq. (\ref{E0t_04}) for ${\tilde E}_0$, I define the following function
\begin{equation}
F(\epsilon,z) = z^\epsilon\left({\partial\over\partial z}\right)^\epsilon{1\over 2 \sinh z},
\label{F}
\end{equation}
and write
\begin{equation}
(-1)^\epsilon \sum_{\ell} [a^2eB(2\ell+1)]^{\epsilon}e^{-eB(2\ell+1)a^2t} = t^{-\epsilon}F(\epsilon,a^2eBt),
\label{sum_1}
\end{equation}
to find
\begin{equation}
{\tilde E}_0={L^2eB\over 4\pi^{3/2} a } 
{(2\epsilon -1)!!\over(2\epsilon)!!}\left(l\over a\right)^{2(\epsilon - 1)}\!\! \sum_{n} \int_0^\infty  t^{-(1/2+\epsilon)}F(\epsilon,a^2eBt)e^{-[k^2_z+m^2]a^2t}dt,
\label{E0t_05}
\end{equation}
where I used $\Gamma(1/2)=\sqrt{\pi}$. This is the exact form of the Lorentz violating correction to the magnetic Casimir energy when $u^\mu$ is parallel to the plates but it can only be reduced to a simple analytic form in three asymptotic cases: $a^{-1}\gg \sqrt{eB} , m$ (small plate distance); $m\gg \sqrt{eB} , a^{-1}$ (large mass); and $\sqrt{eB} \gg m, a^{-1}$ (strong magnetic field). I will examine each of these three cases, and will start with the case of small plate distance.

In the asymptotic case of small plate distance, we have $a^2eB\ll 1$ and $ma\ll 1$. I will take
\begin{equation}
e^{-m^2a^2t}\simeq 1 -m^2a^2t,
\label{exp_1}
\end{equation}
and find, for $z\ll 1$ and $\epsilon $ odd,
\begin{equation}
F(\epsilon,z) \simeq (-1)^\epsilon {\epsilon !\over 2z}+\zeta_R(-\epsilon)(2^\epsilon -1)z^\epsilon,
\label{F2}
\end{equation}
where $\zeta_R$ is the Riemann zeta function, and, for $z\ll 1$ and $\epsilon $ even,
\begin{equation}
F(\epsilon,z) \simeq (-1)^\epsilon {\epsilon !\over 2z}+\zeta_R(-\epsilon-1)(2^{\epsilon+1} -1)z^{\epsilon+1}.
\label{F3}
\end{equation}
Inserting these approximate expansions into Eq. (\ref{E0t_05}), I obtain
\begin{equation}
{\tilde E}_0={L^2\over 4\pi^{3/2} a^3 } 
{(2\epsilon -1)!!\over(2\epsilon)!!}\left(l\over a\right)^{2(\epsilon - 1)}\!\! \sum_{n} \int_0^\infty  t^{-(3/2+\epsilon)}e^{-k^2_z a^2t}\left[
(-1)^\epsilon {\epsilon !\over 2}(1-m^2a^2t)+\zeta_R(-\epsilon)(2^\epsilon -1)(eBa^2t)^{\epsilon+1}\right]dt,
\label{E0t_06}
\end{equation}
for odd values of $\epsilon$, and 
\begin{equation}
{\tilde E}_0={L^2\over 4\pi^{3/2} a^3 } 
{(2\epsilon -1)!!\over(2\epsilon)!!}\left(l\over a\right)^{2(\epsilon - 1)}\!\! \sum_{n} \int_0^\infty  t^{-(3/2+\epsilon)}e^{-k^2_z a^2t}\left[
(-1)^\epsilon {\epsilon !\over 2}(1-m^2a^2t)+\zeta_R(-\epsilon-1)(2^{\epsilon +1} -1)(eBa^2t)^{\epsilon+2}\right]dt,
\label{E0t_07}
\end{equation}
for even values of $\epsilon$. I change the integration variable to $s=k^2_za^2t$ and do the integration to obtain
\begin{eqnarray}
{\tilde E}_0&=&{L^2\over 4\pi a^3 } 
{(2\epsilon -1)!!\over(2\epsilon)!!}\left(l\over a\right)^{2(\epsilon - 1)}\!\! \sum_{n} (k_za)^{2\epsilon+1}\nonumber \\
&\times&\left[
{(-1)^\epsilon\over 2\sqrt{\pi}} {\epsilon !}\left(\Gamma(-\epsilon-{1\over 2})-m^2k_z^{-2}\Gamma(-\epsilon+{1\over 2})\right)+\zeta_R(-\epsilon)(2^\epsilon -1)(eB)^{\epsilon+1}k_z^{-2\epsilon -2}\right],
\label{E0t_08}
\end{eqnarray}
for odd values of $\epsilon$, and 
\begin{eqnarray}
{\tilde E}_0&=&{L^2\over 8\pi a^3 } 
{(2\epsilon -1)!!\over(2\epsilon)!!}\left(l\over a\right)^{2(\epsilon - 1)}\!\! \sum_{n} (k_za)^{2\epsilon+1}
\nonumber \\
&\times&\left[
{(-1)^\epsilon\over \sqrt{\pi}} {\epsilon !}\left(\Gamma(-\epsilon-{1\over 2})-m^2k_z^{-2}\Gamma(-\epsilon+{1\over 2})\right)+\zeta_R(-\epsilon-1)(2^{\epsilon+1} -1)(eB)^{\epsilon+2}k_z^{-2\epsilon -4}\right],
\label{E0t_09}
\end{eqnarray}
for even values of $\epsilon$.

In the case of Dirichlet boundary conditions, I use Eq. (\ref{kz_D}) for $k_z$ and evaluate the infinite sums in terms of the Riemann zeta function $\zeta_R(z)$
\begin{equation}
\sum_{n=1} ^\infty (k_za)^{2\epsilon+1}=\pi^{2\epsilon+1}\zeta_R(-2\epsilon -1),
\label{s_1}
\end{equation}
\begin{equation}
\sum_{n=1} ^\infty (k_za)^{2\epsilon-1}=\pi^{2\epsilon-1}\zeta_R(1-2\epsilon ),
\label{s_2}
\end{equation}
\begin{equation}
\sum_{n=1} ^\infty(k_za)^{-1}=\pi^{-1}\zeta_R( 1),
\label{s_3}
\end{equation}
\begin{equation}
\sum_{n=1} ^\infty (k_za)^{-3}=\pi^{-3}\zeta_R(3),
\label{s_4}
\end{equation}
and obtain ${\tilde E}_0={\tilde E}_0^0+{\tilde E}_0^B$, where ${\tilde E}_0^0$ is the part independent of $B$ and it is the same for $\epsilon$ even and odd,
\begin{equation}
{\tilde E}_0^0=-{L^2\pi^{2\epsilon}\over 4 a^3 } 
\left(l\over a\right)^{2(\epsilon - 1)}
\left[{1\over 2\epsilon +1}\zeta_R(-2\epsilon -1)+{m^2a^2\over 2\pi^2}\zeta_R(1-2\epsilon )\right].
\label{E0t_10}
\end{equation}
Notice that, in the last equation, I used the value of the gamma function for negative semi-integer argument. ${\tilde E}_0^B$ is the $B$-dependent part of ${\tilde E}_0$, given by
\begin{equation}
{\tilde E}_0^B={L^2\over 4\pi^2 a^3 } 
{(2\epsilon -1)!!\over(2\epsilon)!!}\left(l\over a\right)^{2(\epsilon - 1)}
\zeta_R(-\epsilon)\zeta_R(1)(2^\epsilon -1)(eBa^2)^{\epsilon+1},
\label{E0t_11}
\end{equation}
for odd values of $\epsilon$, and 
\begin{equation}
{\tilde E}_0^B={L^2\over 8\pi^4 a^3 } 
{(2\epsilon -1)!!\over(2\epsilon)!!}\left(l\over a\right)^{2(\epsilon - 1)}
\zeta_R(-\epsilon-1)\zeta_R(3)(2^{\epsilon+1} -1)(eBa^2)^{\epsilon+2},
\label{E0t_12}
\end{equation}
for even values of $\epsilon$. I point out that ${\tilde E}_0^0$ of Eq. (\ref{E0t_10}) has a leading term and a mass correction and both are in full agreement with the results of Ref. \cite{ADantas:2023wxd} which are obtained using a different method, the Abel-Plana method. The $B$-dependent part for odd $\epsilon$ , ${\tilde E}_0^B$ of Eq. (\ref{E0t_11}), appears to be divergent since it is proportional to $\zeta_R(1)=\infty$. However this divergence is an artifact of a logarithmic dependence of ${\tilde E}_0^B$ on $a$, as shown in Refs. \cite{Erdas:2023wzy,Erdas:2020ilo,Droguett:2025frq}. ${\tilde E}_0^B$ can be calculated using another method, the generalized zeta function technique discovered by Hawking \cite{Hawking:1976ja}, to find a finite result. I will not repeat here the calculations shown in Refs. \cite{Erdas:2023wzy,Erdas:2020ilo,Droguett:2025frq}, whose outcome is that $\zeta_R(1)$ should be replaced by $\gamma_E+\ln(a\sqrt{eB+m^2})-\ln(2\pi)$, and therefore Eq. (\ref{E0t_11}), valid for odd values of $\epsilon$, becomes
\begin{equation}
{\tilde E}_0^B={L^2\over 4\pi^2 a^3 } 
{(2\epsilon -1)!!\over(2\epsilon)!!}\left(l\over a\right)^{2(\epsilon - 1)}
\zeta_R(-\epsilon)(2^\epsilon -1)(eBa^2)^{\epsilon+1}\left[\gamma_E+\ln\left({a\sqrt{eB+m^2}\over 2\pi}\right)\right],
\label{E0t_13}
\end{equation}
where $\gamma_E=0.577216$ is the Euler-Mascheroni constant.

Moving on to mixed boundary conditions in the small plate distance limit, I use Eq. (\ref{kz_mixed}) for $k_z$ and evaluate the infinite sums in terms of the Hurwitz zeta function $\zeta_H(z,s)=\sum\limits_{\substack{n =0}}^\infty (n+s)^{-z}$
\begin{equation}
\sum_{n=0} ^\infty (k_za)^{2\epsilon+1}=\pi^{2\epsilon+1}\zeta_H(-2\epsilon -1,{1\over 2}),
\label{s_5}
\end{equation}
\begin{equation}
\sum_{n=0} ^\infty (k_za)^{2\epsilon-1}=\pi^{2\epsilon-1}\zeta_H(-2\epsilon +1,{1\over 2}),
\label{s_6}
\end{equation}
\begin{equation}
\sum_{n=0} ^\infty(k_za)^{-1}=\pi^{-1}\zeta_H(1,{1\over 2}),
\label{s_7}
\end{equation}
\begin{equation}
\sum_{n=0} ^\infty (k_za)^{-3}=\pi^{-3}\zeta_H(3,{1\over 2}),
\label{s_8}
\end{equation}
to obtain 
\begin{equation}
{\tilde E}_0^0={L^2\pi^{2\epsilon}\over 4 a^3 } 
\left(l\over a\right)^{2(\epsilon - 1)}
\left[{1\over 2\epsilon +1}(1-2^{-2\epsilon -1})\zeta_R(-2\epsilon -1)+{m^2a^2\over 2\pi^2}(1-2^{1-2\epsilon })\zeta_R(1-2\epsilon )\right],
\label{E0t_14}
\end{equation}
\begin{equation}
{\tilde E}_0^B={L^2\over 4\pi^2 a^3 } 
{(2\epsilon -1)!!\over(2\epsilon)!!}\left(l\over a\right)^{2(\epsilon - 1)}
\zeta_R(-\epsilon)(2^\epsilon -1)(eBa^2)^{\epsilon+1}\left[\gamma_E+\ln\left({a\sqrt{eB+m^2}\over 2\pi}\right)\right],
\label{E0t_15}
\end{equation}
for odd values of $\epsilon$, and 
\begin{equation}
{\tilde E}_0^B={7L^2\over 8\pi^4 a^3 } 
{(2\epsilon -1)!!\over(2\epsilon)!!}\left(l\over a\right)^{2(\epsilon - 1)}
\zeta_R(-\epsilon-1)\zeta_R(3)(2^{\epsilon+1} -1)(eBa^2)^{\epsilon+2},
\label{E0t_16}
\end{equation}
for even values of $\epsilon$. In the last three equations, I used
\begin{equation}
\zeta_H(z,{1\over 2})=(2^z-1)\zeta_R(z),
\label{zh_1}
\end{equation}
and replaced the divergent $\zeta_R(1)$ as I describe in the paragraph above. Notice that the leading term of ${\tilde E}_0^0$ and its mass correction, shown in Eq. (\ref{E0t_14}), are both in full agreement with the results of Ref. \cite{ADantas:2023wxd}, obtained using the Abel-Plana method.

Next, I investigate the asymptotic case of large magnetic field, $a^2eB\gg 1$ and $\sqrt{eB}\gg m$. I first examine Dirichlet boundary conditions, where $k_z$ takes the values shown by Eq. (\ref{kz_D}), and start from the exact Lorentz violating correction of Eq. (\ref{E0t_05}). I do a Poisson resummation of the $n$-summation,
\begin{equation}
\sum_{n=1}^\infty e^{-k^2_z a^2 t}=-{1\over 2}+{1\over 2\sqrt{\pi t}}+{1\over \sqrt{\pi t}}\sum_{n=1}^\infty e^{-n^2/t},
\label{Poisson_1}
\end{equation}
and substitute it into Eq. (\ref{E0t_05}), discarding the first two terms because they lead to the vacuum energy in the presence of one plate only, and to the vacuum energy without boundaries, respectively. Since $a^2eB\gg 1$, I can use an asymptotic expansion of the function $F(\epsilon,z)$ defined in Eq. (\ref{F})
\begin{equation}
F(\epsilon,z) = z^\epsilon\left({\partial\over\partial z}\right)^\epsilon{1\over 2 \sinh z}\simeq z^\epsilon\left({\partial\over\partial z}\right)^\epsilon{e^{-z}}=(-1)^\epsilon z^\epsilon{e^{-z}},
\label{F4}
\end{equation}
and Eq. (\ref{E0t_05}) becomes
\begin{equation}
{\tilde E}_0={L^2(eBa^2)^{\epsilon + 1}\over 4\pi^{2} a^3 } (-1)^\epsilon
{(2\epsilon -1)!!\over(2\epsilon)!!}\left(l\over a\right)^{2(\epsilon - 1)}\!\! \sum_{n=1} ^\infty\int_0^\infty  t^{-1}e^{-n^2/t}e^{-(eB+m^2)a^2t}dt.
\label{E0t_17}
\end{equation}
I change integration variable from $t$ to $s$, with $t={ns\over\sqrt{eB+m^2}a}$ and obtain
\begin{equation}
{\tilde E}_0={L^2(eBa^2)^{\epsilon + 1}\over 4\pi^{2} a^3 } (-1)^\epsilon
{(2\epsilon -1)!!\over(2\epsilon)!!}\left(l\over a\right)^{2(\epsilon - 1)}\!\! \sum_{n=1} ^\infty\int_0^\infty  s^{-1}e^{-na\sqrt{eB+m^2}(s+1/s)}ds,
\label{E0t_18}
\end{equation}
where all summation terms with $n>1$ are negligible because $eBa^2\gg1$. I integrate using
the saddle point method and find
\begin{equation}
{\tilde E}_0={L^2(eBa^2)^{\epsilon + 1}\over 4\pi^{3/2} a^3 } (-1)^\epsilon
{(2\epsilon -1)!!\over(2\epsilon)!!}\left(l\over a\right)^{2(\epsilon - 1)}{e^{-2a\sqrt{eB+m^2}}\over\sqrt{a\sqrt{eB+m^2}}},
\label{E0t_19}
\end{equation}
where we see that the dominant term is the exponential suppression term.

In the case of strong magnetic field and mixed boundary conditions, $k_z$ takes the values shown by Eq. (\ref{kz_mixed}) and I use the following Poisson resummation
\begin{equation}
\sum_{n=0}^\infty e^{-k^2_z a^2 t}={1\over 2\sqrt{\pi t}}+{1\over \sqrt{\pi t}}\sum_{n=1}^\infty (-1)^ne^{-n^2/t},
\label{Poisson_2}
\end{equation}
the asymptotic expansion of Eq. (\ref{F4}), discard the first term in the Poisson resummation because it produces the vacuum energy without boundaries, and follow the same steps I used for the Dirichlet case. Since the only difference is an extra factor of $(-1)^n$, I obtain
\begin{equation}
{\tilde E}_0=-{L^2(eBa^2)^{\epsilon + 1}\over 4\pi^{3/2} a^3 } (-1)^{\epsilon}
{(2\epsilon -1)!!\over(2\epsilon)!!}\left(l\over a\right)^{2(\epsilon - 1)}{e^{-2a\sqrt{eB+m^2}}\over\sqrt{a\sqrt{eB+m^2}}}.
\label{E0t_20}
\end{equation}

Last, I examine the large mass limit, $ma\gg1$ and $m\gg \sqrt{eB}$, and begin by investigating the case of Dirichlet boundary conditions. I use the Poisson resummation of Eq. (\ref{Poisson_1}) into Eq. (\ref{E0t_05}) and drop the first two terms, as I did for the case of strong magnetic field, to obtain
\begin{equation}
{\tilde E}_0={L^2eB\over 4\pi^{2} a } 
{(2\epsilon -1)!!\over(2\epsilon)!!}\left(l\over a\right)^{2(\epsilon - 1)}\!\! \sum_{n=1} ^\infty\int_0^\infty  t^{-(1+\epsilon)}e^{-n^2/t}e^{-m^2a^2t}F(\epsilon, a^2eBt)dt.
\label{E0t_21}
\end{equation}
I change integration variable from $t$ to $s$, with $t={ns\over ma}$ and find
\begin{equation}
{\tilde E}_0={L^2eB\over 4\pi^{2} a } 
{(2\epsilon -1)!!\over(2\epsilon)!!}\left(l\over a\right)^{2(\epsilon - 1)}\!\! \sum_{n=1} ^\infty\left({ma\over n}\right)^\epsilon\int_0^\infty  s^{-(1+\epsilon)}e^{-nma(s+1/s)}F\left(\epsilon, {aeB\over m}ns\right)ds.
\label{E0t_22}
\end{equation}
At this point I neglect all summation terms with $n>1$, since $ma\gg 1$, do the integral using the saddle point method and obtain
\begin{equation}
{\tilde E}_0={L^2eB\over 4\pi^{3/2} a } 
{(2\epsilon -1)!!\over(2\epsilon)!!}\left(l\over a\right)^{2(\epsilon - 1)}\!\! \left({ma}\right)^{(\epsilon-1/2)} F\left(\epsilon, {aeB\over m}\right)e^{-2ma},
\label{E0t_23}
\end{equation}
where, once again, the dominant term is the exponential suppression term. If the dimensionless parameter ${aeB\over m}\ll 1$, I use the asymptotic expansions of $F(\epsilon, z)$ from Eqs. (\ref{F2}) and (\ref{F3}), and find
\begin{equation}
{\tilde E}_0={L^2\over 4\pi^{3/2} a^3 } 
{(2\epsilon -1)!!\over(2\epsilon)!!}\left(l\over a\right)^{2(\epsilon - 1)}\!\! \left({ma}\right)^{(\epsilon+1/2)} \left[(-1)^\epsilon {\epsilon !\over 2}+\zeta_R(-\epsilon)(2^\epsilon -1)({eBa\over m})^{\epsilon+1},\right]e^{-2ma},
\label{E0t_24}
\end{equation}
for odd values of $\epsilon$, and 
\begin{equation}
{\tilde E}_0={L^2\over 4\pi^{3/2} a^3 } 
{(2\epsilon -1)!!\over(2\epsilon)!!}\left(l\over a\right)^{2(\epsilon - 1)}\!\! \left({ma}\right)^{(\epsilon+1/2)} \left[(-1)^\epsilon {\epsilon !\over 2}+\zeta_R(-\epsilon-1)(2^{\epsilon+1} -1)({eBa\over m})^{\epsilon+2},\right]e^{-2ma},
\label{E0t_25}
\end{equation}
for even values of $\epsilon$. Notice that, when $ma\gg1$ and ${aeB\over m}\ll 1$, the leading term of ${\tilde E}_0$ is the same for even and odd $\epsilon$, is independent of $B$ and agrees with the result of Ref. \cite{ADantas:2023wxd}. The magnetic correction, however, is different for odd and even $\epsilon$.

The expression for ${\tilde E}_0$ in large mass approximation with mixed boundary conditions, is obtained using the Poisson resummation of Eq. (\ref{Poisson_2}). Following the same steps I used for the case of Dirichlet boundary conditions, I find that the only difference between the two cases is an extra factor of $(-1)^n$, which leads to an extra factor of $-1$ in the final result, so
\begin{equation}
{\tilde E}_0=-{L^2eB\over 4\pi^{3/2} a } 
{(2\epsilon -1)!!\over(2\epsilon)!!}\left(l\over a\right)^{2(\epsilon - 1)}\!\! \left({ma}\right)^{(\epsilon-1/2)} F\left(\epsilon, {aeB\over m}\right)e^{-2ma},
\label{E0t_26}
\end{equation}
for mixed boundary conditions. When ${aeB\over m}\ll 1$, the value of ${\tilde E}_0$ for large mass and mixed boundary conditions is similar to the two values of Eqs. (\ref{E0t_24}) and (\ref{E0t_25}), but has an extra factor of $-1$. Also in this case the leading term is in full agreement with Ref. \cite{ADantas:2023wxd}.
%%%%%%%%%%%%%%%%%%%%%%%%%%%%%%%%%%%%%%%%%%%%%%%%%%%%%%%%%%%%%%%%%%%%%
\section{Casimir energy for $u^\mu$ perpendicular to the plates}
\label{4}
I use the identity of Eq. (\ref{z})
 to rewrite Eq. (\ref{EB_2}) as
$
E_0=E_0'+{\tilde E}_0,
$
where $E_0'$, the magnetic Casimir energy in absence of Lorentz violation, is given by Eq. (\ref{E0p_01})
and ${\tilde E}_0$, the Lorentz violating correction to the magnetic Casimir energy, is
\begin{equation}
{\tilde E}_0={L^2eB\over 8\pi^{3/2} a } 
{(-1)^\epsilon}\left(l\over a\right)^{2(\epsilon - 1)}\!\! \sum_{n} (ak_z)^{2\epsilon}\int_0^\infty  t^{-1/2}{e^{-(k^2_z+m^2)a^2t}\over\sinh(a^2eBt)}dt.
\label{E0t_27}
\end{equation}

I begin by examining the asymptotic case of small plate distance, where $a^2eB\ll1$ and $ma\ll1$. I use Eq. (\ref{exp_1}) and the following power series approximation valid for $z\ll1$
 \begin{equation}
{1\over \sinh z}\simeq {1\over z} -{z\over 6},
\label{sinh_2}
\end{equation}
into Eq. (\ref{E0t_27}), to obtain
\begin{equation}
{\tilde E}_0\simeq{L^2\over 8\pi^{3/2} a^3 } 
{(-1)^\epsilon}\left(l\over a\right)^{2(\epsilon - 1)}\!\! \sum_{n} (ak_z)^{2\epsilon}\int_0^\infty  t^{-3/2}{e^{-a^2k^2_zt}}\left[{1}-m^2a^2t-{(eBa^2t)^2\over 6}\right]dt.
\label{E0t_28}
\end{equation}
Once I do the $t$-integration, I find
\begin{equation}
{\tilde E}_0={L^2\over 8\pi^{3/2} a^3 } 
{(-1)^\epsilon}\left(l\over a\right)^{2(\epsilon - 1)}\!\! \sum_{n} (ak_z)^{2\epsilon+1}\left[\Gamma(-{1\over 2})-m^2k_z^{-2}\Gamma({1\over 2})-{(eB)^2\over 6k_z^4}\Gamma ({3\over 2})\right].
\label{E0t_29}
\end{equation}
I focus first on Dirichlet boundary conditions. With $k_z$ taking the values shown in Eq. (\ref{kz_D}), I evaluate the infinite sum using Eqs. (\ref{s_1} - \ref{s_2}) and the following identity
\begin{equation}
\sum_{n=1} ^\infty (k_za)^{2\epsilon-3}=\pi^{2\epsilon-3}\zeta_R(-2\epsilon +3),
\label{s_9}
\end{equation}
to obtain
\begin{equation}
{\tilde E}_0=-{L^2\over 8 a^3 } 
{(-1)^\epsilon}\left(l\over a\right)^{2(\epsilon - 1)}\pi^{2\epsilon}\left[2\zeta_R(-1-2\epsilon )+{m^2a^2\over \pi^2}\zeta_R(1-2\epsilon )+{(eBa^2)^2\over 12\pi^4}\zeta_R(3-2\epsilon)\right].
\label{E0t_30}
\end{equation}
Once again, the leading term and its mass correction agree with Ref. \cite{ADantas:2023wxd}.

I treat the case of small plate distance and mixed boundary conditions in a similar way. While $k_z$ takes the values shown in Eq. (\ref{kz_mixed}), I use Eqs. (\ref{s_5} - \ref{s_6})  and the following identity
\begin{equation}
\sum_{n=0} ^\infty (k_za)^{2\epsilon-3}=\pi^{2\epsilon-3}\zeta_H(-2\epsilon +3,{1\over 2}),
\label{s_10}
\end{equation}
to find
\begin{equation}
{\tilde E}_0={L^2\over 4 a^3 } 
{(-1)^\epsilon}\left(l\over a\right)^{2(\epsilon - 1)}\pi^{2\epsilon}\left[(1-2^{-1-2\epsilon})\zeta_R(-1-2\epsilon )+{m^2a^2\over 2\pi^2}(1-2^{1-2\epsilon})\zeta_R(1-2\epsilon )+{e^2B^2a^4\over 24\pi^4}(1-2^{3-2\epsilon})\zeta_R(3-2\epsilon)\right],
\label{E0t_31}
\end{equation}
where I use Eq. (\ref{zh_1}) to write $\zeta_H$ in terms of $\zeta_R$. Also in this case, I find that the leading term and its mass correction agree with Ref. \cite{ADantas:2023wxd}.

I investigate next the strong magnetic field limit, $eBa^2\gg 1$ and $\sqrt{eB}\gg m$, and begin by examining it under Dirichlet boundary conditions. I start from Eq. (\ref{E0t_27}), but cannot  use the Poisson resummation in the form of Eq. (\ref{Poisson_1}) because of the presence of an extra factor of $(ak_z)^\epsilon$ in the sum. In order to circumvent this, I introduce a new parameter $b$ such that,
\begin{equation}
\sum_{n=1}^\infty (ak_z)^\epsilon (-1)^\epsilon e^{-bk^2_z a^2 t}=t^{-\epsilon}\left({\partial\over\partial b}\right)^\epsilon \sum_{n=1}^\infty  e^{-bk^2_z a^2 t},
\label{b_1}
\end{equation}
and now, once I take $b=1$ on the left side and on the right side, after taking the derivatives, I have an alternative form of the infinite sum where I can use the Poisson resummation. I do so on the right side of Eq. (\ref{b_1}) and obtain
\begin{equation}
\sum_{n=1}^\infty (ak_z)^\epsilon (-1)^\epsilon e^{-k^2_z a^2 t}=t^{-\epsilon}\left({\partial\over\partial b}\right)^\epsilon \left[{1\over 2\sqrt{\pi bt}}+{1\over \sqrt{\pi b t}}\sum_{n=1}^\infty e^{-n^2/(bt)}\right]_{b=1},
\label{b_2}
\end{equation}
where I will discard the first term because it leads to the vacuum energy without boundaries. I use Eq. (\ref{b_2})
into Eq. (\ref{E0t_27}) and use the following expansion, valid for $z\gg 1$
\begin{equation}
{1\over \sinh z}\simeq 2e^{-z},
\label{sinh2}
\end{equation}
to find
\begin{equation}
{\tilde E}_0={L^2eB\over 4\pi^{2} a }
\left(l\over a\right)^{2(\epsilon - 1)}\!\! \sum_{n=1} ^\infty\left[\left({\partial\over\partial b}\right)^\epsilon \int_0^\infty  t^{-1-\epsilon}{e^{-n^2/(bt)}\over \sqrt{b}}e^{-(eB+m^2)a^2t}dt\right]_{b=1}.
\label{E0t_32}
\end{equation}
Then I change variable of integration to $s={\sqrt{b}\over n}\sqrt{eB+m^2}at$, neglect all summation terms with $n>1$, and integrate using the saddle point method, to obtain
\begin{equation}
{\tilde E}_0={L^2eB\over 4\pi^{3/2} a }
\left(l\over a\right)^{2(\epsilon - 1)}\!\! (\sqrt{eB+m^2}a)^{\epsilon -1/2}\left[\left({\partial\over\partial b}\right)^\epsilon ({\sqrt{b})^{\epsilon -1/2}}e^{-2a\sqrt{(eB+m^2)/b}}\right]_{b=1}.
\label{E0t_33}
\end{equation}
I find that
\begin{equation}
\left[\left({\partial\over\partial b}\right)^\epsilon {(\sqrt{b})^{\epsilon -1/2}}e^{-2x\sqrt{1/b}}\right]_{b=1}\simeq x^\epsilon e^{-2x},
\label{b_3}
\end{equation}
for $x\gg 1$, and therefore
\begin{equation}
{\tilde E}_0={L^2eBa^{2\epsilon-3/2}\over 4\pi^{3/2}  }
\left(l\over a\right)^{2(\epsilon - 1)}\!\! (eB+m^2)^{\epsilon -1/4} e^{-2a\sqrt{eB+m^2}}.
\label{E0t_34}
\end{equation}

In the strong magnetic field approximation under mixed boundary conditions, $k_z$ takes the values shown in Eq. (\ref{kz_mixed}). I proceed as I did above and in the modified Poisson resummation of Eq. (\ref{b_2}) I obtain an extra factor of $(-1)^n$. This is the only difference between the calculation shown above for Dirichlet boundary conditions and the calculation for mixed boundary conditions. The result is thus the same as in Eq. (\ref{E0t_34}), but with an extra factor of $-1$.

Finally, I examine the large mass limit, $ma\gg 1$ and $m\gg \sqrt{eB}$, under Dirichlet boundary conditions. I introduce the parameter $b$ to do a modified Poisson resummation as in Eq. (\ref{b_2}), and find
\begin{equation}
{\tilde E}_0={L^2eB\over 8\pi^{2} a }
\left(l\over a\right)^{2(\epsilon - 1)}\!\! \sum_{n=1} ^\infty\left[\left({\partial\over\partial b}\right)^\epsilon \int_0^\infty  t^{-1-\epsilon}{e^{-n^2/(bt)}\over \sqrt{b}}{e^{-m^2a^2t}\over\sinh (eBa^2t)}dt\right]_{b=1},
\label{E0t_35}
\end{equation}
next, I change variable of integration to $s={ma\sqrt{b}\over n}t$, neglect all summation terms with $n>1$, and integrate using the saddle point method, to find
\begin{equation}
{\tilde E}_0={L^2eB\over 8\pi^{3/2} a }
\left(l\over a\right)^{2(\epsilon - 1)}\!\! (ma)^{\epsilon -1/2}\left[\left({\partial\over\partial b}\right)^\epsilon {(\sqrt{b})^{\epsilon -1/2}e^{-2am/\sqrt{b}}\over\sinh(eBa/m\sqrt{b})}\right]_{b=1}, 
\label{E0t_36}
\end{equation}
and using the following approximation
\begin{equation}
\left[\left({\partial\over\partial b}\right)^\epsilon {(\sqrt{b})^{\epsilon -1/2}e^{-2x/\sqrt{b}}\over\sinh(y/\sqrt{b})}\right]_{b=1}\simeq {x^\epsilon e^{-2x}\over \sinh (y)},
\label{b_4}
\end{equation}
valid for $x\gg 1$ and $x\gg y$, I find
\begin{equation}
{\tilde E}_0={L^2eB\over 8\pi^{3/2} a }
\left(l\over a\right)^{2(\epsilon - 1)}\!\! (ma)^{2\epsilon -1/2}\ {e^{-2am}\over\sinh(eBa/m)},
\label{E0t_37}
\end{equation}
valid when $eBa/m\ll ma$. If it is the case that $eBa/m\ll 1$, then I use the approximation of Eq. (\ref{sinh_2}) and find
\begin{equation}
{\tilde E}_0\simeq{L^2\over 8\pi^{3/2} a^3 }
\left(l\over a\right)^{2(\epsilon - 1)}\!\! (ma)^{2\epsilon +1/2}\ {e^{-2am}}\left(1-{e^2B^2a^2\over 6m^2}\right),
\label{E0t_38}
\end{equation}
where the leading term, independent of $B$, agrees with what is found in Ref. \cite{ADantas:2023wxd} for the large mass limit, and the magnetic correction lowers the Lorentz violating correction to the Casimir energy.

In the large mass limit under mixed boundary conditions, $k_z$ takes the values shown in Eq. (\ref{kz_mixed}). I proceed as I did above and use the modified Poisson resummation of Eq. (\ref{b_2}), which now has an extra factor of $(-1)^n$. This is the only difference between the calculation shown above for Dirichlet boundary conditions, and the calculation for mixed boundary conditions. The result is the same as in Eqs. (\ref{E0t_37}) and (\ref{E0t_38}), but with an extra factor of $-1$.
%%%%%%%%%%%%%%%%%%%%%%%%%%%%%%%%%%%%%%%%%%%%%%%%%%%%%%%%%%%%%%%%%%%
\section{Casimir pressure}
\label{5}
The Casimir pressure is defined as
\begin{equation}
P_C=-{1\over L^2}{\partial E_0 \over \partial a}
\label{PC_01}
\end{equation}
where $E_0$ is the Casimir energy defined in Eq. (\ref{E0_1}). The Casimir pressure without Lorentz violation, $P_C'$, is well known and can be found, for example, in Refs. \cite{CougoPinto:1998td,Erdas:2013jga,Erdas:2013dha,Erdas:2020ilo}. In this section I will focus on the Lorentz violating corrections to the Casimir pressure, ${\tilde P}_C$.

I begin by examining the situation where $u_\mu$ is parallel to the plates and the scalar field satisfies Dirichlet boundary conditions at the plates. The Lorentz violating correction to the Casimir pressure in the small plate distance limit is given by
\begin{eqnarray}
{\tilde P}_C&=&-{\pi^{2\epsilon}\over 4 a^4 } 
\left(l\over a\right)^{2(\epsilon - 1)}
\left\{\zeta_R(-2\epsilon -1)+(2\epsilon -1){m^2a^2\over 2\pi^2}\zeta_R(1-2\epsilon )+{(2\epsilon -1)!!\over(2\epsilon)!!}
\zeta_R(-\epsilon)(2^\epsilon -1)\right.
\nonumber \\
&\times&
\left.\left({eBa^2\over \pi^2}\right)^{\epsilon+1}\left[\gamma_E+\ln\left({a\sqrt{eB+m^2}\over 2\pi}\right)\right]\right\}
\label{PC_02}
\end{eqnarray}
for $\epsilon$ odd, and 
\begin{equation}
{\tilde P}_C=-{\pi^{2\epsilon}\over 4 a^4 } 
\!\left(l\over a\right)^{2(\epsilon - 1)}\!
\left[\zeta_R(-2\epsilon -1)+(2\epsilon -1){m^2a^2\over 2\pi^2}\zeta_R(1-2\epsilon )+{3\over 2}
{(2\epsilon -1)!!\over(2\epsilon)!!}
\zeta_R(-\epsilon-1)\zeta_R(3)(2^{\epsilon+1} -1)\!\left({eBa^2\over \pi^2}\right)^{\epsilon+2}\right]
\label{PC_03}
\end{equation}
for $\epsilon$ even. In the strong magnetic field regime, ${\tilde P}_C$ is given by
\begin{equation}
{\tilde P}_C={(eBa^2)^{\epsilon + 1}\over 2\pi^{3/2} a^4 } (-1)^\epsilon
{(2\epsilon -1)!!\over(2\epsilon)!!}\left(l\over a\right)^{2(\epsilon - 1)}\left(a\sqrt{eB+m^2}\right)^{1/2}{e^{-2a\sqrt{eB+m^2}}},
\label{PC_04}
\end{equation}
where the dominant term is the exponential suppression term, and it is attractive for odd $\epsilon$ and repulsive for even $\epsilon$. When I move to the large mass limit, I find
\begin{equation}
{\tilde P}_C={eB\over 2\pi^{3/2} a^2} 
{(2\epsilon -1)!!\over(2\epsilon)!!}\left(l\over a\right)^{2(\epsilon - 1)}\!\! \left({ma}\right)^{(\epsilon+1/2)} F\left(\epsilon, {aeB\over m}\right)e^{-2ma},
\label{PC_05}
\end{equation}
displaying again a dominant exponential suppression term.

In the scenario where $u_\mu$ is parallel to the plates and the scalar field satisfies mixed boundary conditions at the plates, I find that the Lorentz violating correction to the Casimir pressure in the small plate distance limit is given by
\begin{eqnarray}
{\tilde P}_C&=&{\pi^{2\epsilon}\over 4 a^4 } 
\left(l\over a\right)^{2(\epsilon - 1)}
\left\{(1-2^{-2\epsilon-1})\zeta_R(-2\epsilon -1)+(2\epsilon -1)(1-2^{1-2\epsilon}){m^2a^2\over 2\pi^2}\zeta_R(1-2\epsilon )+{(2\epsilon -1)!!\over(2\epsilon)!!}
\zeta_R(-\epsilon)(2^\epsilon -1)\right.
\nonumber \\
&\times&
\left.\left({eBa^2\over \pi^2}\right)^{\epsilon+1}\left[\gamma_E+\ln\left({a\sqrt{eB+m^2}\over 2\pi}\right)\right]\right\}
\label{PC_06}
\end{eqnarray}
for odd values of $\epsilon$, and 
\begin{eqnarray}
{\tilde P}_C&=&{\pi^{2\epsilon}\over 4 a^4 } 
\left(l\over a\right)^{2(\epsilon - 1)}
\left[(1-2^{-2\epsilon-1})\zeta_R(-2\epsilon -1)+(2\epsilon -1)(1-2^{1-2\epsilon}){m^2a^2\over 2\pi^2}\zeta_R(1-2\epsilon )
+{21\over 2}
{(2\epsilon -1)!!\over(2\epsilon)!!}\right.
\nonumber \\
&\times&
\left.\zeta_R(-\epsilon-1)\zeta_R(3)(2^{\epsilon+1} -1)\left({eBa^2\over \pi^2}\right)^{\epsilon+2}\right]
\label{PC_07}
\end{eqnarray}
for even values of $\epsilon$. The Lorentz violating correction to the Casimir pressure in the strong magnetic field limit under mixed boundary conditions is given by Eq. (\ref{PC_04}) with an extra factor of $-1$, and in the large mass limit is given by Eq. (\ref{PC_05}) also multiplied by $-1$. Notice how all asymptotic limits of ${\tilde P}_C$ under mixed boundary conditions are attractive when the asymptotic limits of ${\tilde P}_C$ under Dirichlet are repulsive, and vice versa.

Last, I examine the situation where $u_\mu$ is perpendicular to the plates. In the case of Dirichlet boundary conditions, I find
\begin{equation}
{\tilde P}_C=-{{(-1)^\epsilon}\over 4 a^4 } 
\left(l\over a\right)^{2(\epsilon - 1)}\pi^{2\epsilon}\left[(2\epsilon+1)\zeta_R(-1-2\epsilon )+(2\epsilon-1){m^2a^2\over 2\pi^2}\zeta_R(1-2\epsilon )+(2\epsilon-3){(eBa^2)^2\over 24\pi^4}\zeta_R(3-2\epsilon)\right],
\label{PC_08}
\end{equation}
in the small plate distance limit, and 
\begin{equation}
{\tilde P}_C={eBa^{2\epsilon-3/2}\over 2\pi^{3/2}  }
\left(l\over a\right)^{2(\epsilon - 1)}\!\! (eB+m^2)^{\epsilon +1/4} e^{-2a\sqrt{eB+m^2}},
\label{PC_09}
\end{equation}
in the strong magnetic field limit, and
\begin{equation}
{\tilde P}_C={eB\over 4\pi^{3/2} a^2 }
\left(l\over a\right)^{2(\epsilon - 1)}\!\! (ma)^{2\epsilon +1/2}\ {e^{-2am}\over\sinh(eBa/m)},
\label{PC_10}
\end{equation}
in the large mass limit. Finally, in the case of mixed boundary conditions, I obtain
\begin{eqnarray}
{\tilde P}_C&=&{{(-1)^\epsilon}\over 4 a^4 } 
\left(l\over a\right)^{2(\epsilon - 1)}\pi^{2\epsilon}\left[(2\epsilon+1)(1-2^{-1-2\epsilon})\zeta_R(-1-2\epsilon )+(2\epsilon-1)(1-2^{1-2\epsilon}){m^2a^2\over 2\pi^2}\right.
\nonumber \\
&\times&
\left.\zeta_R(1-2\epsilon )+(2\epsilon-3)(1-2^{3-2\epsilon}){(eBa^2)^2\over 24\pi^4}\zeta_R(3-2\epsilon)\right],
\label{PC_11}
\end{eqnarray}
in the small plate distance limit. For mixed boundary conditions, ${\tilde P}_C$ in the strong magnetic field limit and in the large mass limit are given by Eqs. (\ref{PC_09}) and (\ref{PC_10}) multiplied by $-1$, respectively.
%%%%%%%%%%%%%%%%%%%%%%%%%%%%%%%%%%%%%%%%%%%%%%%%%%%%%%
\section{Discussion and conclusions}
\label{6}
In this paper I used the zeta function technique to investigate the Casimir effect of a Lorentz-violating scalar field in the presence of a magnetic field. This charged and massive scalar field satisfies a modified Klein-Gordon equation that breaks Lorentz symmetry in a CPT-even aether-like manner, with the breaking implemented by a constant space-like unit four-vector $u^\mu$ coupled to higher order derivatives of the field. I studied the case of this field satisfying Dirichlet and mixed boundary conditions on two flat parallel plates perpendicular to the magnetic field. I did not study Neumann boundary conditions since they produce the same results as Dirichlet boundary conditions.
In Sec. \ref{3}, I obtained simple analytic expressions for the Casimir energy in the asymptotic cases of short plate distance, strong magnetic field and large mass when $u^\mu$ is parallel to the plates, for both Dirichlet and mixed boundary conditions. In Sec. \ref{4} I did the same for the case of  $u^\mu$ perpendicular to the plates.
In Sec. \ref{5} I obtained simple analytic expressions of the Casimir pressure in the three asymptotic cases for both types of boundary conditions and for the two scenarios where $u^\mu$ is parallel and perpendicular to the plates. 

The Casimir pressure is the measurable quantity, so below I list my results for the Lorentz violating correction to the Casimir  pressure, under Dirichlet boundary conditions, for the three asymptotic limits, the two directions of $u^\mu$, and for critical exponent $\epsilon =2,3$.
\begin{itemize}
  \item $\epsilon =2$ and $u^\mu$ parallel
  
  \underline{ Short plate distance}
 \begin{equation}
{\tilde P}_C={\pi^{4}\over 16 a^4 } 
\!\left(l\over a\right)^{2}\!
\left[{1\over 63}-{m^2a^2\over 20\pi^2}-{21\over 160}
\zeta_R(3)\left({eBa^2\over \pi^2}\right)^{4}\right]
\label{end_01}
\end{equation}
where $\zeta_R(3)=1.20206\cdots$,

 \underline{ Strong magnetic field} 
 \begin{equation}
{\tilde P}_C={3a^{5/2}\over 16\pi^{3/2}  } \left(l\over a\right)^{2}(eB)^3\left(eB+m^2\right)^{1/4}{e^{-2a\sqrt{eB+m^2}}}
\label{end_02}
\end{equation}

 \underline{ Large mass}
 \begin{equation}
{\tilde P}_C={3a^{1/2}\over 16\pi^{3/2} } 
\left(l\over a\right)^{2}\!\! {m}^{5/2} eB F\left(2, {aeB\over m}\right)e^{-2ma}
\label{end_03}
\end{equation}
where $F(2,z)={z^2\over 2}\left({\cosh^2z +1\over \sinh^3z}\right)$.

 \item $\epsilon =2$ and $u^\mu$ perpendicular
  
 \underline{ Short plate distance}
 \begin{equation}
{\tilde P}_C={\pi^{4}\over 16 a^4 } 
\left(l\over a\right)^{2}\left[{5\over 63}-{m^2a^2\over 20\pi^2}+{(eBa^2)^2\over 72\pi^4}\right]
\label{end_04}
\end{equation}

 \underline{ Strong magnetic field}
 \begin{equation}
{\tilde P}_C={a^{5/2}\over 2\pi^{3/2}  }
\left(l\over a\right)^{2}eB(eB+m^2)^{9/4} e^{-2a\sqrt{eB+m^2}}
\label{end_05}
\end{equation}

  \underline{ Large mass}
  \begin{equation}
{\tilde P}_C={a^{5/2}\over 4\pi^{3/2} }
\left(l\over a\right)^{2}m^{9/2}\ {eB\over\sinh(eBa/m)}e^{-2am}.
\label{end_06}
\end{equation}

 \item $\epsilon =3$ and $u^\mu$ parallel
  
  \underline{ Short plate distance}
  \begin{equation}
{\tilde P}_C=-{\pi^{6}\over 96 a^4 } 
\!\left(l\over a\right)^{4}\!
\left\{{1\over 10}-{5m^2a^2\over 21\pi^2}+{7\over 16}
\left({eBa^2\over \pi^2}\right)^{4}\left[\gamma_E+\ln\left({a\sqrt{eB+m^2}\over 2\pi}\right)\right]\right\}
\label{end_07}
\end{equation}

  \underline{ Strong magnetic field}
  \begin{equation}
{\tilde P}_C=-{5a^{9/2}\over 32\pi^{3/2} } \left(l\over a\right)^{4}(eB)^4\left(eB+m^2\right)^{1/4}{e^{-2a\sqrt{eB+m^2}}}
\label{end_08}
\end{equation}

  \underline{ Large mass}
    \begin{equation}
{\tilde P}_C={5\over 32\pi^{3/2} a^{1/2} } \left(l\over a\right)^{4}m^{7/2}eB F\left(3,{aeB\over m}\right){e^{-2ma}}
\label{end_09}
\end{equation}
where $F(3,z)=-{z^3\over 2}\left({\cosh^2z+5\over\sinh^3z}\right)\coth z$.

 \item $\epsilon =3$ and $u^\mu$ perpendicular
 
  \underline{ Short plate distance}
 \begin{equation}
{\tilde P}_C={\pi^{6}\over 96 a^4 } 
\left(l\over a\right)^{4}\left[{7\over 10}-{5m^2a^2\over 21\pi^2}+{(eBa^2)^2\over 40\pi^4}\right]
\label{end_10}
\end{equation}

  \underline{ Strong magnetic field}
 \begin{equation}
{\tilde P}_C={a^{9/2}\over 2\pi^{3/2}  }
\left(l\over a\right)^{4}eB(eB+m^2)^{13/4} e^{-2a\sqrt{eB+m^2}}
\label{end_11}
\end{equation}

  \underline{ Large mass}
  \begin{equation}
{\tilde P}_C={a^{9/2}\over 4\pi^{3/2}  }
\left(l\over a\right)^{4}m^{13/2} {eB\over\sinh(eBa/m)}e^{-2am}.
\label{end_12}
\end{equation}
\end{itemize}

The Casimir pressure without Lorentz violation, $P'_C$, under Dirichlet boundary conditions and for the three asymptotic limits, is listed below

  \underline{ Short plate distance}
 \begin{equation}
P'_C=-{\pi^{2}\over 48 a^4 } 
\left\{{1\over 5}-{m^2a^2\over \pi^2}+{(eBa^2)^2\over \pi^4}\left[\gamma_E+\ln\left({a\sqrt{eB+m^2}\over 2\pi}\right)\right]\right\}
\label{end_13}
\end{equation}

  \underline{ Strong magnetic field}
 \begin{equation}
P'_C=-{eB\over 2\pi^{3/2}a^{1/2}  }
(eB+m^2)^{3/4} e^{-2a\sqrt{eB+m^2}}
\label{end_14}
\end{equation}

  \underline{ Large mass}
  \begin{equation}
P'_C=-{m^{3/2}\over 4\pi^{3/2} a^{1/2} }
 {eB\over\sinh(eBa/m)}e^{-2am}.
\label{end_15}
\end{equation}

Notice that, while the Casimir pressure without Lorentz violation is attractive in all three asymptotic limits, the Lorentz violating correction for $\epsilon = 2$ weakens the attractive Casimir pressure. The $\epsilon = 2 $ magnetic part of the Lorentz violating correction is attractive in the short plate distance limit, when $u^\mu$ is parallel to the plates, and therefore it increases the attractive Casimir pressure while, in all other cases, the $\epsilon = 2 $ magnetic part of the Lorentz violating correction weakens the attractive Casimir pressure. When $\epsilon=3$, the Lorentz violating correction and its magnetic part increase the attractive Casimir pressure when  $u^\mu$ is parallel to the plates, while they lower the  attractive Casimir pressure when  $u^\mu$ is perpendicular to the plates. I find that what I observe for $\epsilon =2$ is true for all even values of $\epsilon$, and what I observe for $\epsilon = 3$ is true for all odd values of $\epsilon$.

Finally, a brief comment on the Casimir pressure under mixed boundary conditions. Everything is reversed in that case, when it is compared to the case of Dirichlet boundary conditions. The Casimir pressure without Lorentz violation is repulsive in all three asymptotic limits, and the Lorentz violating corrections are mostly attractive. With the exception of minor differences in the numerical coefficients in the short plate distance limit, the Lorentz violating correction to the Casimir pressure is the opposite of the Lorentz violating correction under Dirichlet boundary conditions.
%%%%%%%%%%%%%%%%%%%%%%%%%%%%%%%%%%%%%%%%%%%%%%%%%%%%%%%%%%%%%%%%%%%

%%%%%%%%%%%%%%%%%%%%%%%%%%%%%%%%%%%%%%%%%%%%%%%%%%%%%%%%%%%%%%%%%%%
\end{document}